\documentclass[aps,prl,reprint,superscriptaddress,nofootinbib,floatfix]{revtex4-2}

\usepackage{newtxtext}
\usepackage{amsmath,amssymb,bm,mathtools}
\usepackage{newtxmath}
\usepackage{graphicx}
\usepackage{microtype}
\usepackage{tikz}
\usepackage{xcolor}

\let\oldparagraph\paragraph
\renewcommand{\paragraph}{\medskip\oldparagraph}

\usepackage[pdftex]{hyperref}
\definecolor{apsblue}{HTML}{2E3092}
\hypersetup{
    colorlinks=true,
    linkcolor=apsblue,
    citecolor=apsblue,
    urlcolor=apsblue
}

\newcommand{\Tr}{\operatorname{Tr}}
\newcommand{\one}{\openone}
\newcommand{\xv}{\bm{x}}
\newcommand{\EE}{\mathcal{E}}
\newcommand{\EG}{\mathcal{E}_{\rm G}}
\newcommand{\DNG}{\Delta_{\rm NG}}
\newcommand{\ketbra}[1]{|#1\rangle\!\langle #1|}

\usepackage{orcidlink}

\begin{document}

\title{Bosonic Williamson Majorization and Second Laws for Non-Gaussian Work}

\author{Kangze Li\orcidlink{0000-0001-5258-1167}}
\affiliation{College of Physics and Optoelectronic Engineering and Shanxi Key Laboratory of Precision Measurement Physics, Taiyuan University of Technology, Taiyuan 030024, China}
\affiliation{School of Mathematical Sciences and Centre for the Mathematics and Theoretical Physics of Quantum Non-Equilibrium Systems, University of Nottingham, University Park, Nottingham NG7 2RD, United Kingdom}

\author{Leah Turner\orcidlink{0009-0000-4471-2155}}
\affiliation{School of Mathematical Sciences and Centre for the Mathematics and Theoretical Physics of Quantum Non-Equilibrium Systems, University of Nottingham, University Park, Nottingham NG7 2RD, United Kingdom}

\author{Francesco Anna Mele\orcidlink{0000-0002-7900-5167}}
\affiliation{Institute for Quantum Information and Matter, California Institute of Technology,
Pasadena, California 91125, USA}

\author{Gerardo Adesso\orcidlink{0000-0001-7136-3755}}
\email{gerardo.adesso@nottingham.ac.uk}
\affiliation{School of Mathematical Sciences and Centre for the Mathematics and Theoretical Physics of Quantum Non-Equilibrium Systems, University of Nottingham, University Park, Nottingham NG7 2RD, United Kingdom}
\affiliation{Algorithmiq S.r.L., Via della Chiusa 15, 20123 Milano, Italy}

%\date{\today}

\begin{abstract}
How much work a quantum state can deliver depends on the controls available to manipulate it. In bosonic systems, Gaussian operations can leave useful work locked in non-Gaussian states. We show that this locked work has a mode-resolved structure, and we prove a majorization law governing how Gaussian operations can redistribute it on average: locked work can be concentrated into fewer modes, but never evened out. For pure states, this constraint is equivalent to a family of second laws for non-Gaussian work, one for every positive quadratic Hamiltonian. 
These results have sharp consequences for Fock-state batteries. 
Two battery states holding identical energy and identical locked work can be impossible to convert into one another by Gaussian operations, even approximately within the same photon-number sector. Adding a non-Gaussian element such as a Kerr nonlinearity restores deterministic conversion. 
\end{abstract}

\maketitle

\paragraph{\textit{Introduction.}---}
Quantum thermodynamics asks not only how much energy a system contains, but how much of that energy can actually be turned into useful work. The answer depends on the operations available in an experiment. With unrestricted cyclic control, in which a time-dependent perturbation returns the Hamiltonian to its initial form and can therefore implement any unitary, the \emph{ergotropy} is the maximal work that can be extracted from the state, given by the largest achievable decrease of its mean energy~\cite{Allahverdyan2004}. If the controls are restricted, some of this work can remain inaccessible~\cite{Pusz1978,HorodeckiOppenheim2013,Brandao2015,HokkyoUeda2026,MaityHamazaki2026,WatanabeTakagi2026,Guarnieri2026,PoloRodriguez2026}.
%Thus the same state may be more or less useful depending on the physical toolbox used to manipulate it.

Bosonic systems provide a particularly important example. Linear optical transformations, squeezing and displacements are generated by Hamiltonians at most quadratic in the canonical quadratures; together with Gaussian noise and Gaussian measurements, they form the \emph{Gaussian toolbox}~\cite{Weedbrook2012,Lami2018}. These operations are central to quantum optics and continuous-variable technologies, but are not universal~\cite{Walschaers2021}. Restricting control to Gaussian processes is known to have sharp consequences: 
it limits how much work can be
extracted, how efficiently and precisely batteries can be charged~\cite{Brown2016,FriisHuber2018,Singh2019,Singh2023,Cavaliere2025,Andolina2025}, and how
effectively thermal machines can cool a system~\cite{Serafini2020,Narasimhachar2021,Yadin2022,MartinezPaz2013,FreitasPaz2017}.

The energetic cost of this constraint has been quantified~\cite{Brown2016,Singh2019,Hovhannisyan2020,QuintelaRodriguez2025,Jafar2026}. For positive quadratic Hamiltonians, the \emph{Gaussian ergotropy} is defined as the largest decrease of the mean energy achievable with Gaussian unitaries alone~\cite{Brown2016,Hovhannisyan2020,QuintelaRodriguez2025}.
Its difference from the ordinary ergotropy, the \emph{non-Gaussian work potential}, answers a scalar question: how much useful work remains locked unless non-Gaussian control is available?

A single number, however, cannot tell how this locked work is distributed among the modes.
%However, these scalar quantities cannot characterize how the locked work is distributed among the modes. 
For instance, two states may lock the same total work, one in a single mode and the other spread evenly over several, yet they may not be interconvertible by Gaussian operations. The difference matters whenever energy must be delivered in a prescribed form, as in isoergotropic transformations, which rearrange useful energy at fixed total value~\cite{Malavazi2026}, and in \emph{quantum batteries}, whose performance depends on how precisely energy is stored~\cite{FriisHuber2018,Ferraro2018,Rossini2020,Campaioli2024}. The extreme example is a Fock-state battery, whose stored energy exhibits no fluctuations at all, rendering it attractive for powering quantum computation~\cite{Kurman2026}.

In this Letter, we establish a fundamental law governing how Gaussian operations can reshape the distribution of locked work. We show that locked work is encoded in a single, Hamiltonian-independent \emph{Williamson profile}, which lists how strongly each normal mode of a state is excited above the vacuum. We show that Gaussian operations, including measurements and feed-forward, can concentrate this profile into fewer modes, but cannot spread it out more evenly on average. For pure states, this law is equivalent to a whole family of \emph{second laws}: for every positive quadratic Hamiltonian, the non-Gaussian work potential cannot increase on average under Gaussian operations. At fixed particle number, where the profile becomes the spectrum of natural occupations, the same law forbids Gaussian operations from fragmenting a Bose--Einstein condensate~\cite{Mueller2006}.

Our analysis yields a sharp no-go result for batteries. Two Fock-state batteries can store the same energy and the same locked work, one with its photons unevenly split between two modes and the other balanced, yet Gaussian operations cannot turn the first into the second. Even approximate conversion at fixed photon number fails: its success probability vanishes as the battery grows, and for the battery state $|N-1,1\rangle$ so does the best attainable fidelity. A single nonlinear element, such as a Kerr interaction, makes the conversion deterministic, and our framework quantifies the non-Gaussian resources it requires.

Our results unify and extend several previously disconnected lines of work. The symplectic rank~\cite{AdessoSolo,Mele2026} counts how many modes carry non-Gaussianity in the second moments, with implications for Gaussian compression and complexity~\cite{Mele2026}; our Williamson profile refines it by also recording how much locked work each mode carries. For fermions, Williamson occupations underlie computable non-Gaussianity monotones~\cite{Tarabunga2026} and a majorization theory~\cite{Turkeshi2026}; bosons call for a different theory, since their occupations are unbounded and can be concentrated by Gaussian postselection. As a by-product, our law generates strong non-Gaussianity monotones, including the convex roof of the relative entropy of non-Gaussianity~\cite{Albarelli2018,Park2019}. It thus governs at once the resource theory of bosonic non-Gaussianity and its thermodynamic applications.

\paragraph{\textit{Williamson profile.}---}
%Gaussian unitaries act linearly on the first and second moments of a state, so the covariance matrix is the natural place to look for the structure of locked work.
Consider $n$ bosonic modes with canonical quadratures $\hat{\bm R}=(\hat q_1,\hat p_1,\ldots,\hat q_n,\hat p_n)^T$, obeying $[\hat R_j,\hat R_k]=i(\Omega_n)_{jk}$, where $\Omega_\ell=\bigoplus_{j=1}^\ell\bigl(\begin{smallmatrix}0&1\\-1&0\end{smallmatrix}\bigr)$ is the symplectic form on $\ell$ modes. A finite-energy pure state $|\psi\rangle$, with density operator $\psi = \ketbra{\psi}$, has covariance matrix $(V_\psi)_{jk}=\langle\{\Delta\hat R_j,\Delta\hat R_k\}\rangle$, normalized so that the vacuum has covariance $\one$ ($\hbar=1$). By Williamson's theorem~\cite{Williamson1936,Weedbrook2012}, a symplectic transformation, implemented physically by a Gaussian unitary, transforms any such matrix to the normal form
\begin{equation}
V_\psi\longmapsto\bigoplus_{j=1}^{n}\nu_j\one_2,
\qquad 1\le\nu_1\le\cdots\le\nu_n ,
\label{eq:williamson}
\end{equation}
which decomposes the fluctuations of the state into $n$ normal modes, uncorrelated at the level of second moments.

For pure states $\psi$, all $\nu_j=1$ if and only if the state is Gaussian~\cite{Mele2026}. Therefore, each $\nu_j>1$ signals second moments that no pure Gaussian state can display, and the number of such eigenvalues is the \emph{symplectic rank} $s(\psi)$~\cite{AdessoSolo,Mele2026}. This rank quantifies how many non-Gaussian normal modes are present, but not how strongly each is excited. This information is carried by the Williamson occupation $x_j(\psi)=[\nu_j(\psi)-1]/2$, the mean number of excitations in the $j^{\text{th}}$ normal mode. We collect these occupations in the \emph{Williamson profile} 
\begin{equation}
\xv(\psi)=\big(x_1(\psi),\ldots,x_n(\psi)\big),\qquad 0\le x_1\le\cdots\le x_n ,
\label{eq:profile}
\end{equation}
and denote its lower-tail sums by $X_k(\psi)=\sum_{j=1}^{k}x_j(\psi)$. A pure Gaussian state has $\xv=\boldsymbol{0}$, and for a general pure state $s(\psi)$ is simply the number of nonzero entries of $\xv(\psi)$.

The profile has a direct physical interpretation. By Williamson's theorem, the second moments of a pure non-Gaussian state amount to those of a product of independent thermal modes with effective occupations $x_j$. Gaussian unitaries may change which modes carry these fluctuations, but not the ordered values $x_j$. The lower-tail sum $X_k(\psi)$ then answers a concrete question: after the best possible Gaussian change of mode basis, how little excitation can $k$ canonical modes carry? 
To make this precise for an arbitrary covariance matrix $V$, let $X_k(V)=\tfrac12\sum_{j=1}^k[\nu_j^\uparrow(V)-1]$, so that $X_k(V_\psi)=X_k(\psi)$. The symplectic Ky Fan principle~\cite{BhatiaJain2015,DelaRosaPotot2026} gives
\begin{equation}
X_k(V)=\frac14\min_{\substack{S\in\mathbb R^{2n\times2k}\\ S^T\Omega_n S=\Omega_k}}\Tr(S^TVS)-\frac{k}{2},
\label{eq:variational}
\end{equation}
where the columns of $S$ define $k$ canonical modes and $\tfrac14\Tr(S^TVS)-\tfrac{k}{2}$ is their total mean excitation. For $k=n$, the full sum $X_n$ will turn out to be proportional to the total non-Gaussian work potential of equal-frequency oscillators.

To compare the lower-tail sums of two states with different numbers
of modes, we can pad the shorter Williamson profile with zeros by
appending vacuum modes to the state with fewer modes. Alternatively, we can make the same comparison without appending any modes by using the \emph{capped sums}
\begin{equation}
M_t(\psi)=\sum_j\min\{x_j(\psi),t\},\qquad t>0.
\label{eq:capped}
\end{equation}
Each mode contributes its occupation up to the cap $t$, so appending a vacuum mode leaves $M_t$ unchanged. In fact, the inequality $M_t(\psi)\ge M_t(\phi)$ holds for every $t>0$ if and only if $X_k(\psi)\ge X_k(\phi)$ for every $k$ after padding the profiles to the same length~\cite{MarshallOlkin}. Moreover, the capped sums are additive under tensor products, $M_t(\psi\otimes\phi)=M_t(\psi)+M_t(\phi)$. Finally, the function $t\mapsto M_t(\psi)$ determines the positive Williamson occupations and their multiplicities, since its slope
drops precisely at those occupation values and the size of each
drop equals the number of modes sharing that value.

\paragraph{\textit{Williamson majorization and non-Gaussianity monotones.}---}
While Gaussian unitaries leave the profile invariant, a Gaussian
instrument --- implemented %in finitely many steps 
using Gaussian ancillas, unitaries, measurements, and classical
feed-forward~\cite{GiedkeCirac2002,Weedbrook2012,Lami2018} ---
can reshape it by conditioning on measurement outcomes. Our central result is the following monotonicity law:

\emph{\textbf{Theorem 1 (Bosonic Williamson majorization).}
If a Gaussian instrument takes a finite-energy pure state $\psi$ to
pure conditional states $\phi_\alpha$ with outcome distribution
$p(d\alpha)$, then
\begin{equation}
X_k(\psi)\ge \int X_k(\phi_\alpha)\,p(d\alpha)
\qquad \text{for every } k,
\label{eq:main}
\end{equation}
where vacuum modes are added to states with fewer modes so that all
profiles have the same length.}

In words, Gaussian protocols may, on average, \emph{concentrate} the non-Gaussianity visible in second moments into fewer modes, but cannot spread it out more evenly. For example, Hong--Ou--Mandel interference on a balanced beam splitter sends $|1,1\rangle$ to $(|2,0\rangle-|0,2\rangle)/\sqrt2$, and heterodyne detection with outcome $\alpha$ on one output leaves the other in $|\phi_\alpha\rangle\propto |2\rangle-\alpha^{*2}|0\rangle/\sqrt2$. After padding with a vacuum mode, the input profile $(1,1)$ becomes $(0,x_\alpha)$. For outcomes sufficiently close to $\alpha=0$, which occur with nonzero probability, $x_\alpha>1$. Thus, the largest occupation has increased from $1$ to $x_\alpha$, yet the lower-tail sums have not: $X_1$ decreases from $1$ to $0$, while $X_2$ decreases from $2$ to $x_\alpha\le2$. This illustrates why Theorem~1 constrains the lower-tail sums $X_k$ on average, without requiring every individual occupation $x_k$ to decrease.

%Thus the largest occupation need not decrease. What cannot increase on average is the sum of the $k$ smallest occupations.

The proof is given in the End Matter. It compares the average covariance after a Gaussian measurement with the covariance obtained by measuring the Gaussian state that has the same second moments. Concavity of the sums of the smallest symplectic eigenvalues~\cite{BhatiaJain2015,DelaRosaPotot2026} turns this comparison into the lower-tail bound, and a Gaussian purification closes the argument. The bound is necessarily collective: individual conditional symplectic eigenvalues need not decrease.

It follows that the lower-tail sums $X_k$ are only the simplest members of a whole family of \emph{strong non-Gaussianity monotones}, quantities that cannot increase on average under Gaussian instruments~\cite{TakagiZhuang2018,Albarelli2018,Park2019,Lami2018,TurnerGutaAdesso2025}. The capped sums of Eq.~\eqref{eq:capped} are (additive) strong monotones too, being minima of monotones, $M_t=\min_{0\le k\le L}\{X_k+(L-k)t\}$ with $X_0=0$. More generally, any continuous, nondecreasing, concave function $f$ on $[0,\infty)$ with $f(0)=0$ is a nonnegative combination of $x$ and the caps $\min\{x,t\}$, so $\sum_j f(x_j)$ is a nonnegative combination of $X_L$ and the $M_t$, and hence a strong monotone. For instance, $\sum_j x_j^q$, with $0<q\le1$, interpolates between the total occupation $X_L$ at $q=1$ and the symplectic rank $s$ as $q\to0$, while $f(x)=(x+1)\ln(x+1)-x\ln x$ gives the relative entropy of non-Gaussianity of pure states~\cite{GenoniParis2010,MarianMarian2013}. Likewise, for pure states every R\'enyi entropy of the Gaussian state with the same covariance matrix yields a strong monotone; at order two this is $\tfrac12\ln\det V_\psi$, the Gaussian R\'enyi-2 entropy~\cite{Adesso2012Renyi}. For mixed states, we extend $X_k$, $M_t$ and $\sum_jf(x_j)$ by convex roofs, i.e., as minimal averages over pure-state decompositions; these extensions are strong monotones as well [End Matter]. This includes the convex roof of the relative entropy of non-Gaussianity~\cite{Albarelli2018}, whose strong monotonicity also follows from Ref.~\cite{Park2019}.

The criteria in Theorem~1 can also detect no-go results that several well-known measures of non-Gaussianity cannot. For example, it rules out the deterministic Gaussian conversion $|1,4\rangle\to|2,2\rangle$, since the smallest Williamson occupation would increase from $1$ to $2$. Yet the relative entropy of non-Gaussianity~\cite{GenoniParis2010,MarianMarian2013}, Wigner logarithmic negativity~\cite{Albarelli2018}, and stellar rank~\cite{Chabaud2020,HahnConversion2026} all decrease, while the symplectic rank~\cite{Mele2026} remains unchanged.
%Weighted combinations of these monotones will turn out to provide ``second laws'' for non-Gaussian work.

Theorem~1 parallels Nielsen's majorization criterion for entanglement~\cite{Nielsen1999} and its ensemble form, in which tail sums of the smallest Schmidt coefficients cannot increase on average~\cite{Vidal1999,JonathanPlenio1999}; a similar criterion  also governs pure Gaussian entanglement manipulation~\cite{Giedke2003}. Here Williamson occupations take the place of Schmidt coefficients, but they are unbounded and Eq.~\eqref{eq:main} is necessary rather than sufficient. Indeed, the conversion $|1\rangle\to(2|0\rangle+|2\rangle)/\sqrt5$ satisfies Eq.~\eqref{eq:main}, since the input and target have Williamson occupations $1$ and $1/5$, respectively, yet is impossible even with postselection, as it would increase the stellar rank from one to two~\cite{Chabaud2020,HahnConversion2026}. Previous weak-supermajorization results for symplectic spectra concern matrix orbits and pinchings~\cite{HuangMishra2024,TemjensangbaPaul2026,TemjensangbaMishra2026}; to our knowledge, Eq.~\eqref{eq:main} is the first such law for state conversion under bosonic Gaussian protocols, a counterpart of the fermionic theory of Ref.~\cite{Turkeshi2026}.

\paragraph{\textit{Second laws for non-Gaussian work.}---}
The Williamson profile acquires a direct thermodynamic meaning once a Hamiltonian is specified. Consider a positive quadratic Hamiltonian
\begin{equation}
 \hat H=\frac12(\hat{\bm R}-\bm d)^T h(\hat{\bm R}-\bm d),\qquad h>0,
\end{equation}
where $\bm d\in\mathbb R^{2n}$ is a displacement vector and $h$ is real symmetric and positive definite, with symplectic frequencies $\omega_1\le\cdots\le\omega_n$. For this Hamiltonian we denote the ergotropy by $\EE^{(H)}$, the Gaussian ergotropy by $\EG^{(H)}$, and the non-Gaussian work potential by $\DNG^{(H)}=\EE^{(H)}-\EG^{(H)}$~\cite{QuintelaRodriguez2025}. For a pure state, the latter is fixed entirely by the Williamson profile:
\begin{align}
 \DNG^{(H)}(\psi)&=\sum_{j=1}^n w_jx_j(\psi),\nonumber\\
 w_j&=\omega_{n+1-j},\qquad w_1\ge\cdots\ge w_n>0.
\label{eq:workweighted}
\end{align}
The reversed order of the weights reflects how Gaussian unitaries lower the energy: they pair the smallest occupation with the largest frequency, the next smallest with the next largest, and so on~\cite{QuintelaRodriguez2025}. Using $x_j=X_j-X_{j-1}$ and $X_0=0$, Eq.~\eqref{eq:workweighted} can be recast as a combination of lower-tail sums,
\begin{equation}
\DNG^{(H)}=w_nX_n+\sum_{k=1}^{n-1}(w_k-w_{k+1})X_k,
\label{eq:abel}
\end{equation}
with nonnegative coefficients. On a common padded set of modes, with $\hat H$ fixed on that set, Theorem~1 therefore implies that every Gaussian protocol $\psi\xrightarrow{\rm G}\{p(d\alpha),\phi_\alpha\}$ obeys
\begin{equation}
\DNG^{(H)}(\psi)\ge\int\DNG^{(H)}(\phi_\alpha)\,p(d\alpha)
\quad\text{for all positive quadratic }\hat H .
\label{eq:secondlaws}
\end{equation}
Conversely, frequency spectra approaching a step function isolate each $X_k$ [End Matter], so Theorem~1 is equivalent to this whole family of inequalities. The Williamson profile is the single, Hamiltonian-independent object that generates them.

We call Eqs.~\eqref{eq:secondlaws}  the \emph{second laws} for non-Gaussian work, by analogy with the family of monotones providing second laws of quantum thermodynamics~\cite{Brandao2015}: on average, Gaussian processing cannot increase the work that only non-Gaussian control can unlock. Since Gaussian instruments may exchange energy with ancillas and measuring devices, these laws constrain the work locked in the system, not the energy balance of the protocol. 
%Thus energy is a scalar, but under Gaussian control non-Gaussian work has a spectrum: different Hamiltonians weigh this spectrum differently, and the symplectic rank \cite{Mele2026} merely counts its nonzero entries.

\paragraph{\textit{Constraints on condensate fragmentation.}---}
For states with a fixed number of particles, the Williamson profile simplifies. Let $|\psi\rangle$ contain exactly $N$ bosons, with one-body density matrix $\Gamma_{ij}=\langle \hat a_i^\dagger \hat a_j\rangle$. A fixed particle number forces all first moments and all anomalous moments $\langle \hat a_i\hat a_j\rangle$ to vanish, so the passive mode transformation that diagonalizes $\Gamma$ also brings the covariance matrix to Williamson form, with
\begin{equation}
\nu_j=2\lambda_j^\uparrow(\Gamma)+1,\qquad x_j=\lambda_j^\uparrow(\Gamma),
\label{eq:1rdm}
\end{equation}
where $\lambda_j^\uparrow(\Gamma)$ are the eigenvalues of $\Gamma$ in nondecreasing order. The Williamson profile is thus exactly the spectrum of \emph{natural occupations}, the standard diagnostic that distinguishes a simple Bose--Einstein condensate, with a single macroscopically occupied orbital, from a fragmented one with several~\cite{Mueller2006}.

In this setting Theorem~1 becomes a many-body statement. If a Gaussian protocol deterministically converts one pure $N$-boson state into another, then both profiles sum to $N$, and the lower-tail inequalities \eqref{eq:main} reduce to ordinary majorization,
\begin{equation}
 \lambda(\Gamma_{\rm in})\prec \lambda(\Gamma_{\rm out}),
\label{eq:frag}
\end{equation}
where $a\prec b$ means $\sum_{j=1}^k a_j^\downarrow\le\sum_{j=1}^k b_j^\downarrow$ for $k\le n$. Gaussian processing can therefore concentrate the natural occupations but never increase fragmentation, even though particle number need not be conserved at intermediate stages. Fragmenting a condensate requires non-Gaussian dynamics, such as strong repulsive interactions in a double-well trap~\cite{Mueller2006}. A similar majorization structure underlies the resource theory of fermionic one-body entanglement~\cite{Gigena2020}.

\paragraph{\textit{Limitations of Fock-state batteries.}---}
Fixed photon number is also the natural setting for Fock-state batteries~\cite{Kurman2026}. Consider equal-frequency oscillators, $\hat H=\omega\sum_j \hat n_j$ with $\hat n_j=\hat a_j^\dagger \hat a_j$ (dropping the vacuum energy). Every pure state with $N$ photons has energy $N\omega$, and by Eq.~\eqref{eq:workweighted} its non-Gaussian work potential is $\omega X_n=N\omega$: all of its energy is extractable, but none of it by Gaussian unitaries. Energy and locked work are thus constant across the photon-number shell, while the profile ranges from $(0,\ldots,0,N)$ to a nearly uniform distribution.

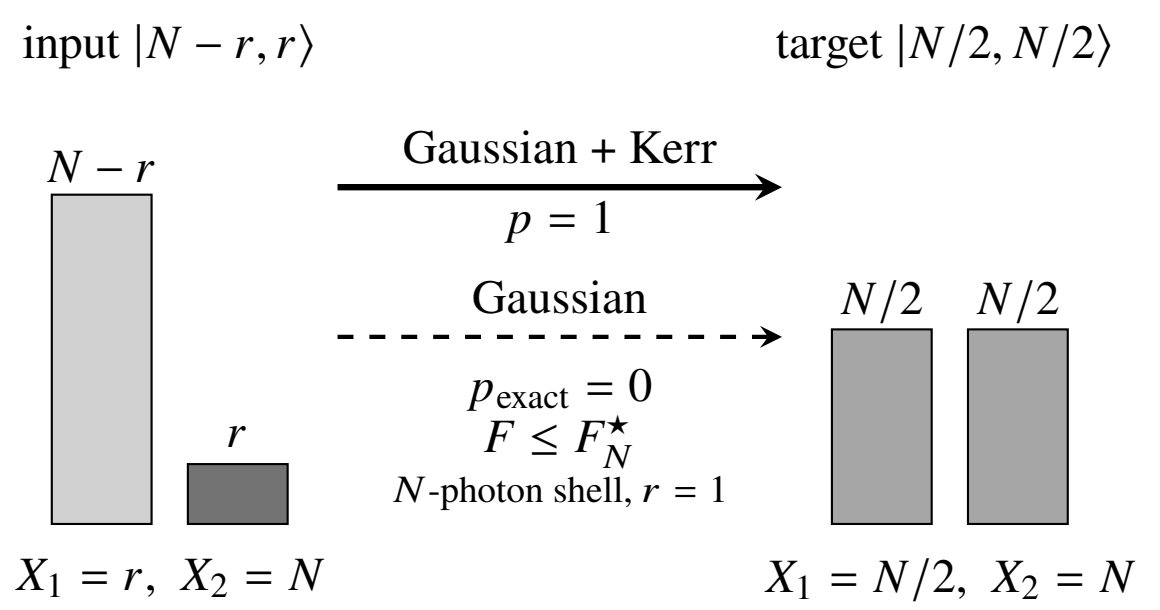
\begin{figure}[t]
\centering
\begin{tikzpicture}[x=0.60cm,y=0.33cm,>=stealth]
  % input battery
  \node[anchor=south] at (0.4,8.8) {input $|N-r,r\rangle$};
  \draw[fill=black!18] (-0.9,0) rectangle (0.2,6.6);
  \draw[fill=black!55] (0.6,0) rectangle (1.7,1.2);
  \node at (-0.35,7.15) {$N-r$};
  \node at (1.15,1.75) {$r$};
  \node[anchor=north] at (0.4,-0.35) {$X_1=r,\;X_2=N$};

  % two distinct conversion routes
  \draw[->,very thick] (2.25,6.75) -- (7.15,6.75);
  \node[align=center] at (4.70,7.55) {Gaussian + Kerr};
  \node[align=center] at (4.70,6.05) {$p=1$};
  \draw[->,dashed,thick] (2.25,3.75) -- (7.15,3.75);
  \node[align=center] at (4.70,4.55) {Gaussian};
  \node[align=center,anchor=north] at (4.70,3.45) {$p_{\rm exact}=0$\\[-1pt] $F\le F_N^\star$\\[-1pt] {\scriptsize $N$-photon shell, $r=1$}};

  % target battery
  \node[anchor=south] at (8.95,8.8) {target $|N/2,N/2\rangle$};
  \draw[fill=black!35] (7.7,0) rectangle (8.8,3.9);
  \draw[fill=black!35] (9.2,0) rectangle (10.3,3.9);
  \node at (8.25,4.45) {$N/2$};
  \node at (9.75,4.45) {$N/2$};
  \node[anchor=north] at (9.0,-0.35) {$X_1=N/2,\;X_2=N$};
\end{tikzpicture}
\caption{\label{fig:battery}Rebalancing a two-mode Fock-state battery, $|N-r,r\rangle\to|N/2,N/2\rangle$ ($N$ even, $1\le r<N/2$); bars indicate the photon number in each mode. Both states have total photon number $N$, non-Gaussian work potential $N\omega$ and symplectic rank two, but the smaller Williamson occupation must rise from $r$ to $N/2$. Gaussian operations cannot reach the target exactly, and within the $N$-photon shell their success probability and fidelity are bounded by Eqs.~\eqref{eq:battery} and~\eqref{eq:shellfid} (the latter for $r=1$). Adding Kerr dynamics to passive linear optics gives full control of the shell and reaches the target deterministically.}
\end{figure}

Take the redistribution task $|N-r,r\rangle\to|\Phi_N\rangle$, with $|\Phi_N\rangle=|N/2,N/2\rangle$, $N$ even and $1\le r<N/2$ (Fig.~\ref{fig:battery}). The two battery states have the same energy, the same locked work, the same symplectic rank ($=2$)~\cite{Mele2026}, and even the same stellar rank ($=N$)~\cite{Chabaud2020}. Only their profiles differ, $(r,N-r)$ versus $(N/2,N/2)$, and this difference turns out to be decisive.

Rebalancing requires the smaller occupation to grow from $r$ to $N/2$, which Theorem~1 forbids for any deterministic Gaussian protocol. The law also bounds probabilistic and approximate rebalancing. If a Gaussian protocol produces, with probability $p_{\rm G}$, an output $\sigma$ in the $N$-photon shell with $\langle\Phi_N|\sigma|\Phi_N\rangle\ge1-\epsilon$, then Eq.~\eqref{eq:main} and a continuity estimate [End Matter] give
\begin{equation}
 X_1(\sigma)\ge \frac N2-N\sqrt\epsilon,
 \qquad
 p_{\rm G}\le\frac{2r}{N(1-2\sqrt\epsilon)}
 \quad(\epsilon<1/4).
\label{eq:battery}
\end{equation}
For fixed $\epsilon<1/4$ and $r$, the success probability vanishes at least as fast as $1/N$: as the batteries grow, the uneven one cannot be rebalanced by Gaussian control, even approximately.%, although both hold exactly the same energy and the same locked work.

The algebraic structure of Fock states sharpens this conclusion further. Writing a pure state through its stellar (Bargmann) function~\cite{ChabaudMehraban2022}, a polynomial times a Gaussian, every Gaussian operation acts on the highest-degree part of the polynomial by a linear change of variables [End Matter]. For Fock states of equal photon number this yields a sharp \emph{merging rule}: a Gaussian instrument maps $|\bm k\rangle$ to $|\bm k'\rangle$, even by postselecting a single measurement outcome, only if the nonzero occupations of $\bm k'$ are obtained by merging those of $\bm k$. Here, merging means partitioning the occupied input modes into groups $B_\ell$, one for each occupied output mode $\ell$, so that $k'_\ell=\sum_{j\in B_\ell} k_j$. Fock occupations can thus merge, as in Hong--Ou--Mandel interference on a balanced beam splitter followed by postselection on the heterodyne outcome $\alpha=0$ at one output, but a single input occupation cannot be divided between output modes, so no Gaussian protocol produces $|\Phi_N\rangle$ from $|N-r,r\rangle$ with nonzero probability.

Within the photon-number shell, which is the natural setting for an energy-preserving redistribution, the same argument also caps the attainable fidelity with the target. For $r=1$, every Gaussian conditional output $\sigma$ supported in the two-mode $N$-photon shell, pure or mixed, obeys [End Matter]
\begin{equation}
 \langle\Phi_N|\sigma|\Phi_N\rangle\le F_N^\star=2^{1-N}\binom{N}{N/2}\Bigl(\frac{N-2}{N-1}\Bigr)^{\frac N2-1}\simeq\sqrt{\frac{8}{\pi eN}},
\label{eq:shellfid}
\end{equation}
so that $F_4^\star=1/2$ and $F_6^\star=2/5$: no Gaussian output in the shell comes close to the target, and the best attainable fidelity vanishes as the battery grows.

Collective processing softens but does not remove the obstruction. Additivity of $M_t$ implies that any deterministic block conversion $\psi^{\otimes n_{\rm in}}\xrightarrow{\rm G}\phi^{\otimes n_{\rm out}}$ satisfies [End Matter]
\begin{equation}
 \frac{n_{\rm out}}{n_{\rm in}}\le \inf_{t:M_t(\phi)>0}\frac{M_t(\psi)}{M_t(\phi)},
\label{eq:rate}
\end{equation}
which for the battery, $|\psi\rangle=|N-r,r\rangle$ and $|\phi\rangle=|\Phi_N\rangle$, reads $n_{\rm out}/n_{\rm in}\le 1/2+r/N$. If the output only reaches global fidelity $1-\epsilon$ within the $(n_{\rm out}N)$-photon shell, the bound relaxes to $n_{\rm out}/n_{\rm in}\le(\frac12+r/N)/(1-\sqrt\epsilon)$. Since such conversions change the photon number, the merging rule does not apply, and the bound follows from Theorem~1 alone.

\paragraph{\textit{Overcoming limitations with non-Gaussian resources.}---}
The identified obstruction to rebalancing batteries is specifically Gaussian, and the Williamson profile also quantifies what it takes to overcome it. Adding a single-mode self-Kerr Hamiltonian $\chi(t)\hat n_1^2$ --- a standard nonlinear ingredient of universal continuous-variable technologies~\cite{LloydBraunstein1999} --- to the passive linear optics toolbox gives full control of the two-mode $N$-photon shell and maps $|N-r,r\rangle$ to $|\Phi_N\rangle$ deterministically [End Matter]. Passive optics cannot change the natural occupations, so the smaller occupation can only grow through the Kerr term, at a rate proportional to $|\chi|$; hence the integrated strength $\int|\chi|\,dt$ is bounded below by the deficit $N/2-r$ divided by the maximal rate [End Matter]. Numerically optimized protocols achieve deterministic rebalancing with total Kerr phase $\int|\chi|\,dt$ between $1$ and $1.25$ for all $N\le20$, essentially independently of $r$.

The advantage persists under photon loss. If both modes lose photons at rate $\kappa$ and the controls conserve photon number, every loss event leaves the $N$-photon shell, so the target probability is the loss-free value times $e^{-N\kappa T}$ [Eq.~\eqref{eq:lossfid}, End Matter]. For $|3,1\rangle\to|2,2\rangle$, an optimized Kerr protocol with total phase $1.25$ thus succeeds with probability $e^{-5\kappa/\chi}$, beating the passive-optics ceiling $3/8$ whenever $\kappa/\chi \lesssim 0.2$.

Alternatively, the non-Gaussian resource may be supplied as a finite-energy ancillary state $\tau$ that is consumed by an otherwise Gaussian protocol, $\ketbra{N-r,r}\otimes\tau\xrightarrow{\rm G}\ketbra{\Phi_N}$. Since the capped sums are additive and cannot increase on average under Gaussian operations, such a deterministic conversion requires $M_{N/2}(\tau)\ge \frac N2-r$ [End Matter].
The ancilla $\tau$ must therefore carry an amount of non-Gaussianity, measured by $M_{N/2}(\tau)$, that grows linearly with $N$ for fixed $r$, even though the battery states $|N-r,r\rangle$ and $|\Phi_N\rangle$ have the same energy and the same symplectic rank. In particular, as $M_t$ is upper bounded by the mean photon number, the ancilla $\tau$ must contain at least $N/2-r$ photons on average, held in non-Gaussian form. A finite-energy catalyst returned exactly at the end of the protocol cannot help either [End Matter].

\paragraph{\textit{Discussion and outlook.}---}
How much work can be drawn from a quantum system depends on the controls at hand. We have shown that, for bosonic systems under Gaussian control, the answer is not a single number: the work that Gaussian operations leave locked is distributed among the modes, and this distribution can be concentrated but never evened out on average. As for entanglement~\cite{Nielsen1999} and thermal operations~\cite{HorodeckiOppenheim2013,Brandao2015}, the constraint takes the form of a majorization law. Its consequences are concrete: two quantum batteries can store the same energy and the same locked work and still be impossible to convert into one another with Gaussian control. Non-Gaussian resources are thus needed not only to extract work, but also to reshape it.

Two questions stand out. First, which constraints, together with the Williamson profile, give necessary and sufficient conditions for converting pure states with Gaussian operations? 
The battery example suggests that a complete criterion must include information beyond the covariance matrix, such as the stellar structure of the state~\cite{Chabaud2020,HahnSimulation2025,HahnConversion2026}. Second, how much nonlinearity is needed to rebalance a battery? We have a lower bound on the Kerr strength and explicit protocols that reach the target; the true minimum lies between the two, and how it changes in the presence of noise remains to be investigated. More broadly, Theorem~1 constrains any process in which Gaussian operations redistribute non-Gaussian excitations among modes, from the preparation of resource states for optical quantum computing~\cite{Walschaers2021} to Gaussian thermal machines~\cite{Serafini2020,Narasimhachar2021} and measurement-assisted protocols~\cite{Kua2025}.

For pure states, homodyne or heterodyne measurements of the covariance matrix determine the Williamson profile without full state tomography~\cite{Laurat2005,DAuria2009,Eichler2011,Lvovsky2009}. Purity must be assessed independently: for mixed states, the convex-roof monotones cannot be inferred from the covariance matrix alone. The four-photon battery provides a direct test through the target population. Passive optics reaches $|2,2\rangle$ from $|3,1\rangle$ with probability at most $3/8$, whereas Kerr-assisted control can exceed this ceiling under the loss model of Eq.~\eqref{eq:lossfid}. Superconducting microwave cavities support Fock-state preparation~\cite{Heeres2015}, native self-Kerr nonlinearities~\cite{Kirchmair2013}, and programmable beam splitters~\cite{Gao2018}, making them a natural platform.

\begin{acknowledgments}
\paragraph{\textit{Acknowledgments.}---}
K.Z.L. gratefully acknowledges support from the National
Natural Science Foundation of China (Grant No.~62405210)
and the Fundamental Research Program of Shanxi Province
(Grant No.~202403021212252). L.T. and G.A. acknowledge financial support from the Engineering and Physical Sciences Research Council  (EPSRC Grants No.~EP/W524402/1 and UKRI6536).

The authors used the large language models ChatGPT (OpenAI) and Claude (Anthropic) as research assistants, for help with constructing and checking proofs, numerical verification, and organizing the literature. The research was conceived and directed by the authors, who developed and verified all results and take full responsibility for the content of this paper.
\end{acknowledgments}

\paragraph{\textit{Data availability.}---}
The data supporting the findings of this study are available from the authors upon reasonable request.

%\clearpage

\section*{END MATTER}

\paragraph{\textit{Proof of Theorem~1.}---}

Gaussian unitaries and pure Gaussian ancillas preserve the profile up to vacuum padding. Mixed Gaussian ancillas and measurement seeds can be purified, while discarded modes can be measured and their outcomes included in the record. It therefore suffices to consider one rank-one Gaussian measurement; finite protocols follow by iteration.

For a pure input $\psi_{AB}$, denote the measured subsystem by $A$ and the retained subsystem by $B$. They contain $a$ and $m$ modes, respectively. Write $V_{AB}=\bigl(\begin{smallmatrix}V_A&C\\ C^T&V_B\end{smallmatrix}\bigr)$ and let $\Sigma$ be the covariance of the pure Gaussian measurement seed. For outcome $\alpha$, the conditional state on $B$ has covariance $V_\alpha$ and mean $\bm\mu_\alpha$. The law of total covariance gives $V_B=\int V_\alpha\,p(d\alpha)+2\operatorname{Cov}(\bm\mu_\alpha)$. The conditional mean is the optimal estimator of the retained quadratures from the measurement record. Its average residual covariance cannot exceed that of the best linear estimator:
\begin{equation}
 \overline V_B=\int V_\alpha\,p(d\alpha)
 \preceq V_B-C^T(V_A+\Sigma)^{-1}C=:V_B^{\rm G}.
\label{eq:mmse}
\end{equation}

Define $F_k(V):=\sum_{j\le k}\nu_j^\uparrow(V)=2X_k(V)+k$.
Take the Gaussian reference state with covariance $V_{AB}$ and purify to a pure Gaussian state on $ABE$, where $E$ has $a+m$ modes. After performing the same measurement $\Sigma$ on $A$, the conditional state on $BE$ is pure Gaussian, with covariances $V_B^{\rm G}$ on $B$ and $V_E^{\rm c}\preceq V_E$ on $E$. Note the two marginals of a pure Gaussian state share their nontrivial symplectic spectra~\cite{Giedke2003}. From this, we can deduce $V_B^{\rm G}$ and $V_E^{\rm c}$ share the same symplectic spectra, up to vacuum padding to match dimensions, hence $F_k(\one_A\oplus V_B^{\rm G})=F_k(V_E^{\rm c})$. Also, by considering the marginals on $AB$ and $E$ of the pure covariance matrix $V_{ABE}$, we see $F_k(V_E)=F_k(V_{AB})$. We may then write
\begin{equation}
\begin{aligned}
 \int F_k(\one_A\oplus V_\alpha)\,p(d\alpha)
 &\le F_k\big(\one_A\oplus\mbox{${\int}$} V_\alpha\,p(d\alpha)\big) \\
 &\le F_k(\one_A\oplus V_B^{\rm G})
 =F_k(V_E^{\rm c}) \\
 &\le F_k(V_E)=F_k(V_{AB}),
\end{aligned}
\label{eq:proofchain}
\end{equation}
where we note from Eq.~\eqref{eq:variational} $F_k(V)$ is a minimum of linear functionals of $V$, hence concave, and is Loewner-monotone, therefore the first inequality follows from Jensen's inequality, the second inequality from Eq.~\eqref{eq:mmse}, and the third inequality from monotonicity of $F_k$.

Eq.~\eqref{eq:proofchain} is Eq.~\eqref{eq:main} with $L=a+m$, concluding the proof.

\paragraph{\textit{Monotones, second laws and mixed states.}---}
With $X_0=0$, $M_t=\min_{0\le k\le L}\{X_k+(L-k)t\}$. Choose a minimizing $k$ for the input and apply Theorem~1; minimizing separately for each output can only lower the average. Thus $M_t$ is a strong monotone. Every continuous, nondecreasing, concave $f$ with $f(0)=0$ obeys
\begin{equation}
 f(x)=f'(\infty)\,x+\int_0^\infty\min\{x,t\}\,d\mu_f(t),
\label{eq:frep}
\end{equation}
where $f'(\infty)=\lim_{x\to\infty}f(x)/x\ge0$ and $d\mu_f=-df'_+\ge0$, with $f'_+$ the right derivative. Equality follows by differentiation and $f(0)=0$. Hence $\sum_jf(x_j)=f'(\infty)X_L+\int M_t\,d\mu_f(t)$ is a strong monotone. The entropy function $f(x)=(x+1)\ln(x+1)-x\ln x$ has $f'(\infty)=0$ and $d\mu_f=dt/[t(t+1)]$. For pure $\psi$, $\sum_jf(x_j)$ is its relative entropy of non-Gaussianity~\cite{GenoniParis2010,MarianMarian2013}, so its strong monotonicity follows.

The R\'enyi-$\alpha$ entropy of the matching Gaussian state is $\sum_jf_\alpha(x_j)$, with $f_\alpha(x)=\ln[(x+1)^\alpha-x^\alpha]/(\alpha-1)$ and $\alpha>0$. For $x>0$, set $r=x/(x+1)$ and $I_b(r)=\int_0^b r^s\,ds$. Then $f_\alpha'=\frac{\alpha}{\alpha-1}(1-r)I_{\alpha-1}/I_\alpha>0$. For $\alpha>1$, the ratio decreases with $r$ as weight shifts from $[0,\alpha-1]$ towards larger $s$. For $0<\alpha<1$, $|I_{\alpha-1}|$ decreases and $I_\alpha$ increases. Since $1-r$ also decreases, $f_\alpha$ is increasing and concave; $\alpha=1$ follows by continuity. At $\alpha=2$, $\sum_jf_2(x_j)=\tfrac12\ln\det V_\psi$~\cite{Adesso2012Renyi}.

Eq.~\eqref{eq:abel} expresses every pure-state work potential as a nonnegative combination of the $X_k$. Conversely, setting $w_j=1$ for $j\le k$ and $w_j=\delta>0$ otherwise gives $\DNG^{(H)}=X_k+\delta(X_n-X_k)$. Taking $\delta\downarrow0$ recovers each lower-tail inequality. If outcomes of total probability $p$ yield a pure target $\phi$, nonnegativity on the remaining outcomes gives $p\le\min_{k:X_k(\phi)>0}X_k(\psi)/X_k(\phi)$.

For a finite-energy mixed state $\rho$, we define $X_k(\rho)=\inf_\mu\int X_k(\psi)\,d\mu(\psi)$ over probability measures $\mu$ on pure states with $\int\psi\,d\mu(\psi)=\rho$~\cite{Albarelli2018,TakagiZhuang2018,Park2019}, and likewise $M_t(\rho)$ and the convex roof of $\sum_jf(x_j)$; for the relative entropy of non-Gaussianity, the latter is $\delta_{\rm CR}$ of Ref.~\cite{Albarelli2018}. These convex roof quantifiers are strong monotones: refine each Gaussian output to pure branches, apply Theorem~1 to every component of a near-optimal decomposition, and coarse-grain. This establishes property 3a conjectured in Ref.~\cite{Albarelli2018} (which also follows by taking the pure-state convex roof of the strongly monotone measure in Ref.~\cite{Park2019}).
Gaussian states are mixtures of pure Gaussian states, so their convex roofs vanish, and if one factor is pure every decomposition factorizes, giving $M_t(\psi\otimes\sigma)=M_t(\psi)+M_t(\sigma)$. Finally, taking $S$ as the identity in Eq.~\eqref{eq:variational} gives $X_L(V)\le\frac14\Tr V-\frac L2\le\langle\hat n\rangle$, so $M_t(\rho)\le\langle\hat n\rangle_\rho$ by linearity of $\langle\hat n\rangle$.

\paragraph{\textit{Natural occupations and approximate rebalancing.}---}
At fixed particle number, $\langle\hat a_i\rangle=\langle\hat a_i\hat a_j\rangle=0$, so passive diagonalization of $\Gamma$ gives $V=\bigoplus_j[2\lambda_j(\Gamma)+1]\one_2$, proving Eq.~\eqref{eq:1rdm}. For positive $\Gamma,\Gamma'$ of equal trace, the variational formulas $X_k(\Gamma)=\min_P\Tr(P\Gamma)$ over rank-$k$ projectors and $M_t(\Gamma)=\Tr\Gamma-\max_{0\le Q\le\one}\Tr[Q(\Gamma-t\one)]$ imply $|X_k(\Gamma)-X_k(\Gamma')|,\,|M_t(\Gamma)-M_t(\Gamma')|\le\frac12\|\Gamma-\Gamma'\|_1$. Up to transposition, $\Gamma/N$ is the one-body reduced state. Contractivity of trace distance, which transposition preserves, therefore gives, for pure $N$-boson states with $\eta=|\langle\psi|\phi\rangle|^2$,
\begin{equation}
 |X_k(\psi)-X_k(\phi)|,\ |M_t(\psi)-M_t(\phi)|\le N\sqrt{1-\eta}.
\label{eq:fixedcontinuity}
\end{equation}
Every pure component of $\sigma$ supported in the shell remains in it. Applying this estimate to each component and averaging gives $X_k(\sigma)\ge X_k(\phi)-N\sqrt\epsilon$ whenever $\langle\phi|\sigma|\phi\rangle\ge1-\epsilon$, and likewise for $M_t$. Here concavity of the square root bounds the averaged error. Taking $\phi=\Phi_N$, $k=1$, and using $r\ge p_{\rm G}X_1(\sigma)$ proves Eq.~\eqref{eq:battery}.

\paragraph{\textit{Merging rule for Fock states.}---}
In the Bargmann representation, creation and annihilation act as $z_j$ and $\partial_j$. For finite stellar rank $d$, write $\mathcal F_\psi=q\,e^\theta$, where $q$ has total degree $d$ and $\theta$ is quadratic~\cite{Chabaud2020,HahnConversion2026}. Track the highest-degree part $q_d$. A Gaussian unitary changes the Gaussian factor and transforms creation operators into affine combinations of $z_j$ and $\partial_j$. After removing the Gaussian factor, derivatives lower the polynomial degree; only multiplication by linear terms contributes at degree $d$. Gaussian projections likewise replace variables by affine means and add covariance terms, which lower the degree by two. This includes homodyne detection, while Gaussian ancillas add no polynomial factors. Iterating, the output polynomial has degree at most $d$ and degree-$d$ part $q_d\circ\Lambda$ for some linear map $\Lambda$.

For Fock input, $q_d\propto\prod_jz_j^{k_j}$. A Fock output with the same photon number has constant Gaussian factor, requiring $\prod_j\ell_j^{k_j}\propto\prod_lz_l^{k'_l}$, where $\ell_j$ are the rows of $\Lambda$. Unique factorization forces each occupied $\ell_j$ to be proportional to one output variable. Thus $k'_l=\sum_{j:\ell_j\propto z_l}k_j$: occupations merge but cannot split. Conversely, conditioning on the heterodyne outcome $\alpha=0$ at the second output of a balanced beam splitter maps $z_1^az_2^b$ to a multiple of $z_1^{a+b}$, realizing each merger on an ideal branch~\cite{Mele2026}. This excludes $|N-r,r\rangle\to|\Phi_N\rangle$ on every conditional branch.

\paragraph{\textit{Fixed-shell fidelity bound.}---}
For input $|N-1,1\rangle$, a pure conditional output of a Gaussian protocol in the $N$-photon shell is proportional to $\ell^{N-1}\ell'$. Choosing orthonormal modes $\hat b\propto\ell$ and $\hat c$, it lies in $\mathcal S_b=\operatorname{span}\{|N,0\rangle_{bc},|N-1,1\rangle_{bc}\}$. Its target fidelity is at most $\|P_{\mathcal S_b}|\Phi_N\rangle\|^2$. For $\hat b=\cos\vartheta\,\hat a_1+e^{i\varphi}\sin\vartheta\,\hat a_2$, the phase drops out and $u=\sin^2(2\vartheta)$ gives
\begin{equation}
 \|P_{\mathcal S_b}|\Phi_N\rangle\|^2=2^{-N}\binom{N}{N/2}\,u^{\frac N2-1}\bigl[N-(N-1)u\bigr].
 \label{eq:shellproj}
\end{equation}
Maximizing gives $u=(N-2)/(N-1)$ and $F_N^\star$; linearity extends the bound to mixed outputs. The normalized maximizing projection has the form $\ell^{N-1}\ell'$. After rescaling, the coefficient matrix of these forms is a contraction and hence a submatrix of a passive interferometer. Vacuum ancillas and vacuum projections on the extra outputs (i.e., conditioning on $\alpha=0$ in each heterodyne measurement) give the conditional state that attains the bound. The selected outcome may have vanishing probability.

\paragraph{\textit{Rates, ancillas and catalysts.}---}
Additivity gives $n_{\rm in}M_t(\psi)\ge n_{\rm out}M_t(\phi)$ for any deterministic conversion $\psi^{\otimes n_{\rm in}}\xrightarrow{\rm G}\phi^{\otimes n_{\rm out}}$, proving Eq.~\eqref{eq:rate}. For the battery, the ratio is minimized at $t=N/2$, where $M_t(\psi)=N/2+r$ and $M_t(\Phi_N)=N$. If the output has fidelity at least $1-\epsilon$ in the $n_{\rm out}N$-photon shell, Eq.~\eqref{eq:fixedcontinuity} gives $M_{N/2}(\sigma)\ge n_{\rm out}N(1-\sqrt\epsilon)$, proving the approximate rate.

A consumed ancilla must obey $M_t(\psi)+M_t(\tau)\ge M_t(\phi)$, giving $M_{N/2}(\tau)\ge N/2-r$. An exact catalyst $\zeta$ on finitely many modes has finite $M_t(\zeta)$, which cancels between input and output. Since the cap inequalities are equivalent to lower-tail order, no Williamson obstruction is removed. A pure target cannot remain correlated with the catalyst, so the distinction between strong and weak catalysts relevant to Gaussian thermal operations~\cite{Yadin2022} does not arise here.

\paragraph{\textit{Kerr control and its cost.}---}
On the two-mode $N$-photon shell, passive optics generates the Schwinger algebra, including $\hat J_+=\hat a_1^\dagger\hat a_2$. Set $\hat D=\hat n_1^2$. In the basis $|j,N-j\rangle$, $\operatorname{ad}_{\hat D}^\ell(\hat J_+)=\sum_{j<N}(2j+1)^\ell c_j|j+1,N-j-1\rangle\langle j,N-j|$, with $c_j=\sqrt{(j+1)(N-j)}\neq0$. Since the gaps $2j+1$ are distinct, linear combinations of these commutators select each neighbouring transition. Together with their adjoints, these transitions generate $\mathfrak{su}(N+1)$, giving full control of the shell.

Passive rotations preserve the eigenvalues of $\Gamma$. In their interaction picture, subtract the passive term $\chi N\hat n_u$ from $\chi\hat n_u^2$, leaving $-\chi\hat n_u\hat n_v$. Its contribution to $\dot\Gamma$ has zero diagonal and off-diagonal element $i\chi\langle\hat Q_u\rangle$, with $\hat Q_u=[\hat n_u^2-N\hat n_u,\hat a_u^\dagger\hat a_v]$ and $v$ orthogonal to $u$. Let $T_N$ be the real symmetric tridiagonal matrix with off-diagonal entries $\frac12(2j+1-N)c_j$. Multiply $\hat Q_u$ by a phase making its expectation real. The resulting Hermitian part differs from $T_N$ only by basis phases, so $|\langle\hat Q_u\rangle|\le\|T_N\|$. Thus the Kerr contribution to $\dot\Gamma$ has norm at most $|\chi|\|T_N\|$. Weyl's inequality bounds the smaller occupation's rate by the same quantity. Integrating over the required increase $N/2-r$ gives $\int|\chi|\,dt\ge(N/2-r)/\|T_N\|$. Here $\|T_4\|=2\sqrt3$ and $\|T_N\|\simeq N^2/4$, by upper bounds on row sums and trial vectors supported near their maximum. Cross-Kerr couplings obey the same bound.

The protocols were found by gradient optimization of alternating passive rotations and Kerr phases, minimizing total phase at fidelity above $1-10^{-6}$. The lower bound decreases as $1/N$, whereas the optimized phase stays of order one for $N\le20$. Whether the minimum cost tends to a nonzero constant remains open.

\paragraph{\textit{Passive benchmark and photon loss.}---}

For passive unitaries alone, the maximum transition probability from $|N-1,1\rangle$ to $|\Phi_N\rangle$ is
\begin{equation}
 F_{N,\mathrm{pass}}=2^{1-N}\binom{N}{N/2}\left(\frac{N-2}{N}\right)^{N/2-1}.
 \label{eq:passivefid}
\end{equation}
We model equal zero-temperature photon loss in the two modes at rate $\kappa$ using $\dot\rho=-i[\hat H(t),\rho]+\kappa\sum_j\mathcal D[\hat a_j]\rho$, where we assume $[\hat H(t),\hat n_1+\hat n_2]=0$, i.e. the Hamiltonian cannot add back any lost photons, so only evolutions with no photon loss can contribute to the target state $|\Phi_N\rangle$.
Conditioned on no photon loss, the evolution is described by $\hat H_{\text{eff}}(t)=\hat H(t)-\frac{i\kappa}{2}N$, hence
\begin{equation}
 \langle\Phi_N|\rho(T)|\Phi_N\rangle
 =e^{-N\kappa T}|\langle\Phi_N|U(T)|N-r,r\rangle|^2,
 \label{eq:lossfid}
\end{equation}
with $U(T)$ the loss-free propagator. In particular, photon loss cannot increase the maximum passive transition probability~\eqref{eq:passivefid}. 

To see the advantage of Kerr interactions, taking $N=4$ we find $F_{4,\mathrm{pass}}=3/8$, yet eight equal Kerr segments ($\hat H_K=\chi\hat n_1^2$, total phase $1.25$) interspersed with passive rotations reach the target state with fidelity above $1-10^{-12}$. With constant $\chi$ and instantaneous passive gates, $T=1.25/\chi$, so the target population $e^{-5\kappa/\chi}$ exceeds $3/8$ for $\kappa/\chi<\ln(8/3)/5\approx 0.2$, in agreement with direct master-equation simulations.

\bibliography{williamson_work}

\end{document}